\documentclass{article}
\pdfoutput=1

\usepackage{amssymb,amsfonts,amsmath}
\usepackage{cite,enumerate,float}
\usepackage{color}
\usepackage{tikz}
\usetikzlibrary{arrows,snakes,backgrounds}

\usepackage{amssymb,amsfonts,amsmath,stmaryrd}
\usepackage{cite,enumerate,float,indentfirst}
\usepackage{color}
\usepackage{tikz}
\usetikzlibrary{arrows,snakes,backgrounds,calc}
\usepackage{ytableau}
\usepackage[vcentermath]{youngtab}

\newcommand\shadetext[2][]{%
  \setbox0=\hbox{{#2}}%
  \tikz[baseline=0]\path [#1] \pgfextra{\rlap{\copy0}} (0,-\dp0) rectangle (\wd0,\ht0);%
}
\newcommand{\rbtext}[1]{\shadetext[left color=blue, right color=red]{\bfseries #1 } }

\def\be{\begin{eqnarray}}
\def\ee{\end{eqnarray}}
\def\nn{\nonumber}

\def\be{\begin{eqnarray}}
\def\ee{\end{eqnarray}}
\def\nn{\nonumber}

\def\p{\partial}

\def\Tr{{\rm Tr}\,}

\definecolor{red}{rgb}{1,0,0}
\definecolor{orange}{rgb}{1,0.5,0}
\definecolor{violet}{rgb}{0.7,0,1}

\hoffset= - 1.0in         

\begin{document}

\begin{center}
\begin{small}
\hfill MIPT/TH-07/23\\
\hfill FIAN/TD-06/23\\
\hfill ITEP/TH-08/23\\
\hfill IITP/TH-06/23\\

\end{small}
\end{center}

\vspace{.5cm}

\begin{center}
\begin{Large}\fontfamily{cess}
\fontsize{15pt}{27pt}
\selectfont
	\textbf{
Many-body integrable systems implied by the WLZZ models}
	\end{Large}
	
\bigskip \bigskip

\begin{large}
A. Mironov$^{b,c,d}$\,\footnote{mironov@lpi.ru; mironov@itep.ru},
A. Morozov$^{a,c,d}$\,\footnote{morozov@itep.ru}
\end{large}

\bigskip

\begin{small}
$^a$ {\it MIPT, Dolgoprudny, 141701, Russia}\\
$^b$ {\it Lebedev Physics Institute, Moscow 119991, Russia}\\
$^c$ {\it Institute for Information Transmission Problems, Moscow 127994, Russia}\\
$^d$ {\it NRC ``Kurchatov Institute'' - ITEP, Moscow 117218, Russia}\\
\end{small}
 \end{center}

\bigskip

\begin{abstract}
We provide some details about the recently discovered integrable systems
implied by commutativity of $W$ operators along the rays on the plane of roots of $w_\infty$-algebra.
The simplest system of this type is the rational Calogero model,
other systems escaped attention in the past.
Existence of these systems is intimately tied to   the
very interesting WLZZ matrix models, which are now under intensive study.
\end{abstract}

\bigskip

\section{Introduction}

The study of $W$-representations \cite{MSh,Al} in \cite{WLZZ1,WLZZ2} let to discovery of a new class of models called WLZZ models
generated by $\hat W$-operators belonging to a commutative subalgebra
of $w_\infty$ algebra.
Given the role of $w_\infty$ algebras in various branches of modern theory,
this is a very important result, which should provide tools for non-perturbative description
of a big variety of theories.
Indeed, the new partition functions are proved to respect the {\it superintegrability} property \cite{siMI},
and appeared to be (skew) hypergeometric $\tau$-functions \cite{ChR1}.
Description in terms of AMM\slash CEO topological recursion \cite{toprec1,toprec2,toprec3,toprec4,toprec5,toprec6} is still at an early stage \cite{MMsc},
but this issue should be also clarified in the near future. Various deformations of the WLZZ models are also straightforward \cite{ChR3,ChR4}.
Far more interesting appeared to be a matrix model representation of the WLZZ models,
it is provided by a new kind of integrable 2-matrix models \cite{ChR2}
which unify the Hermitian and Kontsevich theories.
The underlying theory is currently under thorough investigation by different groups.

In this paper, we want to attract an additional attention to one of the side-branches
outlined in \cite{ChR2}.
Like the ordinary Hermitian matrix models have a relation to Calogero systems \cite{Hikami,AM1,AM2,Wang},
the WLZZ models appear related to an essentially new integrable many-body generalizations of the rational Calogero model,
which we are going to describe here. All these integrable many-body systems are embedded into $w_\infty$-algebra framework, relations of the Calogero systems with $w_\infty$-algebra being known within various frameworks for many years \cite{Hikami,Awata,Horozov1,Horozov2,Wang}.
This still remains a very preliminary description, but we hope that it will make other people
interested.

This short letter is organized as follows. In sec.2, we describe the general construction of commutative WLZZ families of operators in terms of variables $p_k$. Within this approach, one starts with two auxiliary operators and constructs an infinite set of generating operators, each of them being, at the same time, the first Hamiltonian of an infinite commutative family. The two auxiliary operators generally depends on a parameter $\beta$. In sec.3, we reformulate the constructed commutative sets at $\beta=1$ in terms of matrix operators using an $N\times N$ matrix $\Lambda$ that parameterizes the variables $p_k=\Tr\Lambda^k$. The simplest of these commutative sets gives rise to the rational Calogero Hamiltonians at the free fermion point, the eigenvalues of $\Lambda$ being associated with the coordinates of the Calogero system. Note that, in variance with the realization in terms of $p_k$ variables, the commutative families are absolutely explicit in terms of matrix operators. In sec.4, we consider the generic case of $\beta\ne 1$. Then, the commutative sets are reformulated in terms of the Dunkl operator, also in a very explicit form, and the simplest of the sets gives rise to the rational Calogero Hamiltonians at arbitrary coupling. At last, sec.5 contains some concluding comments.

\section{WLZZ families of operators\label{2}}

In \cite{ChR1,ChR2}, we proposed an infinite set of commutative families of operators that are formed by generators of $w_{1+\infty}$ algebra and are related to hypergeometric and skew hypergeometric $\tau$-functions, and with WLZZ matrix models. The families associated with skew hypergeometric $\tau$-functions describe the positive branch of the WLZZ matrix models and are constructed in the following way: one starts from the pair of operators
\be
\hat W_0&=&\dfrac{1}{2} \sum_{a, b=0}\left((a+b) p_a p_b \frac{\partial}{\partial p_{a+b}}+a b p_{a+b} \frac{\partial^2}{\partial p_a \partial p_b}\right)
\ee
\be\label{hF1}
\hat F_1&=&\left[{\p\over\p p_1},\hat W_0\right]=\sum_{b=0} (b+1)p_b \dfrac{\partial}{\partial p_{b+1}}
\ee
the first of them being the cut-and-join operator \cite{GJ,MMN1,MMN2}. Here we put $p_0=N$.

Using these operators, one constructs a set of generating operators
\be
\hat F_m = {\rm ad}_{\hat W_0} \hat F_{m-1} = {\rm ad}_{\hat W_0}^m \hat F_0
\ee
with $\hat F_0 = -\frac{\p}{\p p_1}$,
and any two consecutive such generating operators $\hat F_{m}$, $\hat F_{m+1}$ give rise to an infinite commutative family ${\cal F}^{(m)}$ of operators:
\be
\hat  H_{n}^{(m)} = {1\over (n-1)!}\ {\rm ad}_{\hat F_{m+1}}^{n-1} \hat F_{m}
\ee

Similarly, in order to describe the families associated with hypergeometric $\tau$-functions associated with the negative branch of the WLZZ matrix models, one better uses the variables $t_k=p_k/k$ and starts from the same cut-and-join operator and the operator
\be\label{E1}
\hat E_1&=&[\hat W_0,p_1]=\sum_{b=0} (b+1)t_{b+1} \dfrac{\partial}{\partial t_{b}}
\ee
Now the terms with $t_0$ do not contribute, instead one puts ${\p\over\p t_0}=N$, which is equivalent by rotating all operators with $e^{Nt_0}$: $O\to e^{-Nt_0} Oe^{Nt_0}$.

Using these operators, one constructs a set of generating operators
\be
\hat E_m = {\rm ad}_{\hat W_0} \hat E_{m-1} = {\rm ad}_{\hat W_0}^m \hat E_0
\ee
with $\hat E_0 = p_1$,
and any two consecutive such generating operators $\hat E_{m}$, $\hat E_{m+1}$ give rise to an infinite commutative family ${\cal E}^{(m)}$ of operators:
\be
\hat  H_{-n}^{(m)} = {1\over (n-1)!}\ {\rm ad}_{\hat E_{m+1}}^{n-1} \hat E_{m}
\ee

One can consider the $\beta$-deformation of this scheme that preserves commutativity. To this end, one slightly modifies the auxiliary cut-and-join operator
\be\label{W0b}
\hat W_0&=&\dfrac{1}{2} \sum_{a, b=1}\left(\beta (a+b) p_a p_b \frac{\partial}{\partial p_{a+b}}+a b p_{a+b} \frac{\partial^2}{\partial p_a \partial p_b}\right) + {1-\beta\over 2}\sum_k (k-1)kp_k \dfrac{\partial}{\partial p_k}
+C_N^\beta\sum_k kp_k \dfrac{\partial}{\partial p_k}
\ee
while $\hat F_1$ does not change at all, and $\hat E_1$ becomes
\be\label{E1b}
\hat E_1=\sum_{b=1} bp_{b+1} \dfrac{\partial}{\partial p_{b}}+C_N^\beta\cdot p_1
\ee
and then one applies the scheme above. The coefficient $C_N^\beta$ in these formulas can be chosen arbitrarily. We choose
\be\label{CNb}
C_N^\beta=N\beta-\beta+1
\ee
which makes formulas of sec.4 simpler (our choice differs from the choice of \cite{WLZZ2}).

\section{Commutative family in terms of matrix derivatives}

In this section, we consider only the case of $\beta=1$.

\subsection{Matrix formalism for positive series\label{Matrix}}

Now we give a more detailed and operational description of the Hamiltonians $H_k^{(m)}$. It can be done in terms of an $N\times N$ matrix $\Lambda$ such that $p_k=\Tr\Lambda^k$. In these terms, the operators introduced earlier are\footnote{Hereafter, by the matrix derivative, we imply the derivative w.r.t. matrix elements of the transposed matrix: $\left(\frac{\partial}{\partial \Lambda}\right)_{ij}=\frac{\partial}{\partial \Lambda_{ji}}$.}:
\be\label{W0L}
\hat W_0={1\over 2}:\Tr \left(\Lambda{\p\over\p\Lambda}\right)^2:+N \Tr \left(\Lambda{\p\over\p\Lambda}\right)
\ = {1\over 2}\Tr \left({\p\over\p\Lambda}\Lambda^2{\p\over\p\Lambda}\right)
\ee
\be\label{F1L}
\hat F_1=\Tr {\p\over\p\Lambda}
\ee
and, further,
\be
\hat F_2=[\hat F_1,\hat W_0]=
\Tr \left( {\p\over\p\Lambda}\Lambda{\p\over\p\Lambda}\right)
\ee
so that the commutative family ${\cal F}^{(1)}$ is
\be\label{F1}
\boxed{
\hat H_k^{(1)}=\frac{1}{(k-1)!} {\rm ad}_{\hat F_2}^{k-1}\hat F_1= \Tr {\p^k\over\p\Lambda^k}
}
\ee
Similarly, one obtains
\be
\hat F_3=[\hat F_2,\hat W_0]=\Tr \left({\p\over\p\Lambda}\Lambda{\p\over\p\Lambda}\Lambda{\p\over\p\Lambda}\right)
\ee
and the commutative family ${\cal F}^{(2)}$ is
\be
\hat H_1^{(2)}=\hat F_2
\ee
\vspace{-0.8cm}
\be
\hat H_2^{(2)}=[\hat F_2,\hat F_3]=\Tr \left( {\p\over\p\Lambda}\Lambda{\p\over\p\Lambda}\right)^2
\ee
\vspace{-0.4cm}
\be
[\hat H_1^{(2)},\hat H_2^{(2)}]=0
\ee
and, generally,
\be
\boxed{
\hat H_k^{(2)} = \Tr \left(\frac{\p}{\p \Lambda} \Lambda \frac{\p}{\p \Lambda}\right)^k
}
\ee

\bigskip

In general the $m$-th commuting system with $\left[\hat H_{k}^{(m)},\hat H_{k'}^{(m)}\right]=0$ is
\be\label{19}
\boxed{
\hat H_k^{(m)} = \Tr \left(\hat {\cal O}^{(m)} \right)^k
}
\ee
where $\hat {\cal O}^{(m)} = \frac{\p}{\p \Lambda} \Lambda \frac{\p}{\p \Lambda} \ldots \Lambda \frac{\p}{\p \Lambda}$
with $m$ $\Lambda$-derivatives and $m-1$ $\Lambda$ in between them.
In particular,
\be
\hat F_m = \hat H_1^{(m)}
\ee
and the two recursions are
\be
\hat F_{m+1} = [\hat W_0,\hat F_m],    \ \ \ \ \ \ \ \ \hat H_{k+1}^{(m)} = [\hat H_k^{(m)},\hat F_{m+1}]
\ee

For example,
\be
\hat H_k^{(3)} = \Tr \left( \frac{\p}{\p\Lambda} \Lambda  \frac{\p}{\p\Lambda} \Lambda \frac{\p}{\p\Lambda}
 \right)^k
\nn \\
\hat F_4 = \Tr \left(
\frac{\p}{\p\Lambda} \Lambda  \frac{\p}{\p\Lambda} \Lambda \frac{\p}{\p\Lambda}\Lambda \frac{\p}{\p\Lambda}
 \right) = \hat H_1^{(4)}
\ee
and
\be
\hat H_{k+1}^{(3)} = {1\over k}[\hat H_k^{(3)},\hat  F_4] =  {1\over k}[\hat H_k^{(3)},\hat  H_1^{(4)}]
\ee

\subsection{Matrix formalism for negative series}

In this case, one has to make the substitution $\Lambda\to\Lambda^{-1}$ and consider all the operators sandwiched between $\det^{-N}\Lambda$ and $\det^{N}\Lambda$: $O(\lambda)\to\det^{-N}\Lambda O(\lambda^{-1})\det^{N}\Lambda$. The first step ($\Lambda\to\Lambda^{-1}$) follows the new version of 2210.09993 (yet to appear). It can be equivalently described as coming to the variables $p_k=\Tr\Lambda^{-k}$ instead of $p_k=\Tr\Lambda^{k}$. The second step is necessary to reproduce the condition ${\p\over\p t_0}=N$.

This recipe follows from the simple fact: the described procedure does not change the operator $\hat W_0$ (\ref{W0L}), while making from $\hat F_1$ (\ref{F1L}) the operator $\hat E_1$ (\ref{E1}) being rewritten in terms of $p_k$ variables. All other operators are generated by these two.

As an illustration, let us also remind that, as explained in \cite{MMsc}, the families ${\cal F}^{(1)}$, ${\cal E}^{(1)}$ can be described by the operators
\be
\hat H_{m}^{(1)}=\Tr\left({\p^{m}\over\p\Lambda^{m}}\right)=\sum_{n=1}p_n\widetilde{W}^{(+,m)}_{n+m}
+N\widetilde{W}^{(+,m)}_{m}\nn\\
\hat H_{-m}^{(1)}=\Tr\left({\p^{m}\over\p\Big(\Lambda^{-1}\Big)^m}\right)=\sum_{n=1}p_n\widetilde{W}^{(-,m)}_{n-m}
\ee
with $p_k=\Tr \Lambda^k$. The second expression implies that one can also write
\be
\hat H_{-m}^{(1)}=\Tr\left({\p^{m}\over\p\Lambda^m}\right)=\sum_{n=1}p_n\widetilde{W}^{(-,m)}_{n-m}
\ee
with $p_k=\Tr \Lambda^{-k}$.

\section{Commutative families as integrable systems}

\subsection{Rational Calogero system}

Each commutative family of constructed operators can be naturally associated with an $N$-body integrable system, the Hamiltonians being just these commuting operators. Since our operators are invariant operators on matrices, they can be realized in terms of eigenvalues. The family of Hamiltonians ${\cal F}^{(1)}$ are associated with the rational Calogero system at the free fermion point, and the coordinates of the Calogero particles are just eigenvalues $\lambda_i$ of the matrix $\Lambda$. The Hamiltonians $H_k^{(1)}$ (\ref{F1}) can be rewritten in these terms \cite[Eq.(21)]{MMM}
\be
\hat H_k^{(1)}=\Tr{\p^k\over\p\Lambda^k}=\sum_i\sum_{{I\subset [1,\dots,N]}\atop{|I|=k-1}}\prod_{j\in I}{1\over\lambda_i-\lambda_j}
{\p\over\p\lambda_i}
\ee
where the operator $\hat H_k^{(1)}$ is understood as acting on invariant functions of $\Lambda$. Note that the sum includes the terms with poles at $i=j$, which are resolved by the L'H\^ospital's rule.

In order to leave this free fermion point, one has to consider the $\beta$-deformation of the operators.
In terms of eigenvalues, i.e. with $p_k=\sum_{i=1}^N \lambda_i^k$, and when acting on symmetric functions of
$\lambda_i$,\footnote{Formal subtleties of the procedure can be found in \cite{SV}.}
the auxiliary operators are
\be
\hat W_0&=&{1\over 2}\sum_i\lambda_i^2{\p^2\over\p\lambda_i^2}+\beta\sum_{i\ne j}{\lambda_i^2\over \lambda_i-\lambda_j}{\p\over\p\lambda_i}+\sum_i\lambda_i{\p\over\p\lambda_i}\nn\\
\hat F_1&=&\sum_i{\p\over\p\lambda_i}\nn\\
\hat F_2&=&\sum_i\lambda_i{\p^2\over\p\lambda_i^2}
+2\beta\sum_{i\ne j}{\lambda_i\over \lambda_i-\lambda_j}{\p\over\p\lambda_i}+\sum_i{\p\over\p\lambda_i}
\ee
and the Hamiltonians are
\be\label{Wn}
\hat H_1^{(1)}&=&\hat F_1=\sum_i{\p\over\p\lambda_i}\nn\\
\hat H_2^{(1)}&=&2\beta\sum_{i\ne j}{1\over \lambda_i-\lambda_j}{\p\over\p\lambda_i}+\sum_i{\p^2\over\p\lambda_i^2}\nn\\
\hat H_3^{(1)}&=&3\beta^2\sum_{i\ne j\ne k}{1\over (\lambda_i-\lambda_j)(\lambda_i-\lambda_k)}{\p\over\p\lambda_i}
+3\beta\sum_{i\ne j}{1\over\lambda_i-\lambda_j}{\p^2\over\p\lambda_i^2}+\sum_i{\p^3\over\p\lambda_i^3}\nn\\
\nn\\
\ldots\nn\\
\nn\\
\hat H_n^{(1)}&=&\sum_{k=1}^nC^n_k\beta^{n-k}\sum_i\sum_{{I\subset [1,\dots,N]\backslash i}\atop{|I|=k}}
\prod_{j\in I}{1\over\lambda_i-\lambda_j}{\p^k\over\p\lambda_i^k}
\ee
where $C^n_k$ are the binomial coefficients. This is a set of the (mutually commuting) rational Calogero-Sutherland Hamiltonians.

One can also realize $\hat H_n$ in terms of the Dunkl operators $\hat D_i$:
\be
\hat D_i={\p\over\p\lambda_i}+\beta\sum_{j\ne i}{1\over \lambda_i-\lambda_j}(1-P_{ij})
\ee
where $P_{ij}$ is the operator permuting $i$ and $j$. When acting on symmetric functions of $\lambda_i$,
\be
\hat H_k^{(1)}=\sum_i\hat D_i^k\Big|_{symm}
\ee
where we manifestly indicated projecting onto symmetric functions of $\lambda_i$.

Note that the standard Calogero-Sutherland Hamiltonians are obtained by the rotation:
\be\label{Cal}
\hat H_k^{Cal}=\Delta(\lambda)^{\beta/2}\cdot \sum_i\hat D_i^k\cdot \Delta(\lambda)^{-\beta}
\ee
where $\Delta(\lambda)=\prod_{i< j}(\lambda_i-\lambda_j)$.

\subsection{Higher families}

Surprisingly, the construction of  sec.\ref{Matrix} is almost directly extended to the $\beta\ne 1$ case. That is, one just has to substitute any matrix derivative ${\p\over\p\Lambda}$ with the Dunkl operator $\hat D_i$, any matrix $\Lambda$ with its eigenvalue $\lambda_i$, the trace with the summation over $i$, and to put $N=1$. This would produce the higher commuting families of the Hamiltonians acting on symmetric functions of $\lambda_i$:
\be
\boxed{
\hat H_k^{(m)} = \sum_i \left(\mathfrak{O}^{(k)}_i
\right)^k\Big|_{symm}
}
\ee
where
\be
\mathfrak{O}^{(m)}_i:=\hat D_i\lambda_i\hat D_i\ldots\lambda_i\hat D_i
\ee
and the Dunkl operator $\hat D_i$ is repeated in this expression $m$ times.

For instance,
\vspace{-0.5cm}
\be
\hat H_k^{(2)}=\sum_i\left(\hat D_i\lambda_i\hat D_i\right)^k\Big|_{symm}\nn\\
\hat H_k^{(3)}=\sum_i\left(\hat D_i\lambda_i\hat D_i\lambda_i\hat D_i\right)^k\Big|_{symm}\nn\\
\ldots
\ee
Note that all the Hamiltonians are covariant w.r.t. the scaling: the transformation $\lambda_i\to\alpha\lambda_i$ with some constant $\alpha$ gives rise to $\hat H_k^{(m)}\to\alpha^k \hat H_k^{(m)}$.

\subsection{A new integrable interaction}

The first example of these Hamiltonians is given by the series $\hat H_k^{(2)}$:
\be
\hat H_1^{(2)}=\hat F_2=\sum_i\lambda_i{\p^2\over\p\lambda_i^2}+2\beta\sum_{i\ne j}{\lambda_i\over \lambda_i-\lambda_j}{\p\over\p\lambda_i}+\sum_i{\p\over\p\lambda_i}
\ee
i.e. one can see that this series starts with the second order Hamiltonian with a not that simple quadratic part, and, hence, there is no momentum conservation law in this case.
Instead, this {\it first} Hamiltonian in the $m=2$ series can be compared with the {\it second} Calogero Hamiltonian
(at $m=1$).
The quadratic part in $H_1^{(2)}$ can be simplified by the change of variables $\lambda_i\to\lambda_i=\mu_i^2$ so that the Hamiltonian which is a counterpart of the Calogero Hamiltonian is
\be
\boxed{
\hat H^{(2)}={\cal N}_2^{-1}\cdot 4\hat H_1^{(2)}\cdot {\cal N}_2=\sum_i\left({\p^2\over\p\mu_i^2}+{1\over 4}{1\over
\mu_i^2}\right)+2\beta(\beta-1)\sum_i\lambda_i{\p^2\log\Delta(\lambda)\over\p\lambda_i^2}\Big|_{\lambda_i=\mu_i^2}
}
\label{H21}
\ee
\vspace{-0.3cm}
$$
{\cal N}_2=\Big(\prod_i\mu_i\Big)^{-1/2}\cdot\Delta(\mu^2)^{-\beta}
$$
Thus we get a non-trivial, {\bf still, integrable} $\beta$-deformation of Calogero system, with a somewhat
non-trivial potential different from the standard Calogero one:
\be
\hat H^{Cal}_2={\cal N}^{-1}\!\cdot\!\hat H_2^{(1)}\!\!\cdot {\cal N} = {\cal N}^{-1}\!\cdot\!\!
\left(\sum_i{\p^2\over\p\lambda_i^2}+2\beta\sum_{i\ne j}{1\over \lambda_i-\lambda_j}{\p\over\p\lambda_i}\right)
\!\!\cdot {\cal N}= \sum_i{\p^2\over\p\lambda_i^2}-\beta(\beta-1)\sum_{i\ne j}{1\over(\lambda_i-\lambda_j)^2}
\label{H12}
\ee
\vspace{-0.3cm}
$$
{\cal N}=\Delta(\lambda)^{-\beta}
$$
Note that, in variance with (\ref{H12}), (\ref{H21}) is non-trivial even for $\beta=1$: it describes $N$ non-interacting particles, however, each of them is in the inverse square potential.

For higher series, the very first Hamiltonian $\hat H^{(m)}_1= \hat{ F}_m$ should be similarly compared with the
higher Calogero Hamiltonians $\hat H_m^{(1)}$. For instance,
\be
\hat H^{(3)}_1=\hat F_3&=&\sum_i\lambda_i^2{\p^3\over\p\lambda_i^3}+3\sum_i\lambda_i{\p^2\over\p\lambda_i^2}+
\sum_i{\p\over\p\lambda_i}+3\beta\sum_{i\ne j}{\lambda_i^2\over\lambda_i-\lambda_j}{\p^2\over\p\lambda_i^2}+\nn\\
&+&2\beta(\beta+2)\sum_{i\ne j}{\lambda_i\over\lambda_i-\lambda_j}{\p\over\p\lambda_i}+
3\beta^2\sum_{i\ne j\ne k}{\lambda_i^2\over(\lambda_i-\lambda_j)(\lambda_i-\lambda_k)}{\p\over\p\lambda_i}
\ee
The $m$-th series begins with the Hamiltonian of the $m$-th order in derivatives,
and the substitution making the highest derivative term simple is $\lambda_i\to\lambda_i=\mu_i^m$:
\be
\hat H_1^{(m)}=\sum_i\lambda_i^{m-1}{\p^m\over\p\lambda_i^m}+\ldots
\stackrel{\lambda_i\to\mu_i^m}{\longrightarrow}\sum_i{\p^m\over\p\mu_i^m}+\ldots
\ee
Moreover, one can again remove the next-to-leading derivative by a simple rescaling:
\be
\hat H^{(m)}={\cal N}_m^{-1}\cdot m^m\hat H_1^{(m)}\cdot
{\cal N}_m=\sum_i\left({\p^m\over\p\mu_i^m}+\xi(\mu_1,\ldots,\mu_N){\p^{m-2}\over\p\mu_i^{m-2}}+\ldots\right)\stackrel{\beta\to 1}{=}
\sum_ih^{(m)}(\mu_i)
\ee
\vspace{-0.3cm}
$$
{\cal N}_m=\Big(\prod_i\mu_i\Big)^{-(m-1)/2}\cdot\Delta(\mu^2)^{-\beta}
$$
 Thus, similarly to the $m=2$ case (\ref{H21}), at $\beta=1$ the Hamiltonian $\hat H^{(m)}$ becomes a system of $N$ non-interacting particles in external potentials, each being described by a one-particle Hamiltonian $h^{(m)}(\mu)$. Moreover, this persists for higher Hamiltonians as well, the one particle Hamiltonian in terms of $\lambda$ being
 \be
h_k^{(m)}(\mu)=\left.   (mk)^{mk}\cdot \lambda^{(m-1)/(2m)}\cdot {\p\over\p\lambda}\cdot\left(\lambda{\p\over\p\lambda}\right)^{mk-1}
 \cdot\lambda^{-(m-1)/(2m)} \right|_{\lambda=\mu^{mk}} = \frac{\p^{mk}}{\p \mu^{mk}} + \ldots
 \ee
This follows from the fact that action of the Hamiltonians (\ref{19}) on the Vandermonde determinant $\Delta(\lambda)$ of the eigenvalues of the matrix $\Lambda$ diagonalizes them.

At $\beta\neq 1$ a non-trivial interaction appears. It depends on $m$, but is always integrable.

\subsection{Negative branch of WLZZ}

 Similarly to the $\beta=1$ case realized in terms of the matrices, in order to describe the eigenvalue realization of the Hamiltonians of ${\cal E}^{(m)}$ families, one has to make in all operators of this section the substitution $\lambda_i\to\lambda_i^{-1}$, and replace all the operators $O(\lambda_i)\to \left(\prod_{i=1}^N\lambda_i\right)^{-C_N^\beta}\cdot
O(\lambda_i^{-1})\cdot\left(\prod_{i=1}^N\lambda_i\right)^{C_N^\beta}$ with $C_N^\beta$ from (\ref{CNb}). Indeed, one can check that this works for the two auxiliary operators, i.e. leaves the cut-and-join operator $\hat W_0$ intact while making $\hat E_1$, (\ref{E1}) from $\hat F_1$, (\ref{hF1}).

\section{Conclusion}

The goal of this paper is to attract more attention to a new generalization of the rational Calogero integrable many-body system.
It is in a non-trivial direction: from quadratic (or even linear) to higher order differential operators for the lowest
Hamiltonian $H_1^{(m)}$ of the commutative set.
The families are made from operators lying on arbitrary rays originating at the origin in the
table of generators (roots) of $w_\infty$ algebra.
These integrable systems are intimately related to the very interesting WLZZ models \cite{WLZZ2},
and were actually discovered in \cite{ChR1,ChR2} as a byproduct of the possibility to
represent WLZZ models as {\it matrix} models.


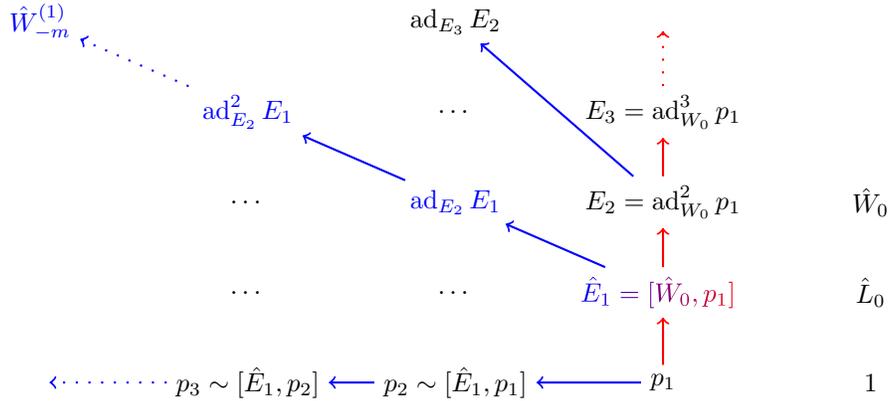
\pgfmathsetmacro{\ra}{2.3}

\begin{center}
\begin{tikzpicture}[scale=1.2]
  \foreach \x [evaluate=\x as \xval using int(\x-1)] in {3,2}
  {
  \node (mx\x) at (-1*\ra*\x,1){ $p_{\x}\sim [\hat E_1,p_{\xval}]$};
  }
  \node (mx1 y0) at (-\ra,1){$p_1$};
    \node (mx4 y0) at (-4*\ra,1) {};
  \node (x0 y1) at (0,1+1) {$\hat L_0$};
  \node (x0 y2) at (0,1+2) {$\hat W_0$};
  \node (mx1 y1) at (-\ra,1+1) {\rbtext{$\hat E_{1}=[\hat W_0,p_1]$}};
  \node (mx2 y1) at (-2*\ra,1+1) {\ldots};
  \node (mx3 y1) at (-3*\ra,1+1) {\ldots};
  \node (mx3 y2) at (-3*\ra,1+2) {\ldots};
  \node (mx2 y3) at (-2*\ra,1+3) {\ldots};
  \node (mx2 y2) at (-2*\ra,1+2) {${\color{blue} \operatorname{ad}_{E_2}E_1 } $ };
\node (mx3 y3) at (-3*\ra,1+3) {$ {\color{blue} \operatorname{ad}^2_{E_2}E_1 }$};
\node (mx1 y2) at (-\ra,1+2) {$E_2=\operatorname{ad}_{W_0}^2 p_1$};
\node (mx1 y3) at (-\ra,1+3) {$E_3=\operatorname{ad}_{W_0}^3 p_1$};
\node (mx1 y4) at (-\ra,1+4) {};
\node (mx2 y4) at (-2*\ra,1+4) {$\operatorname{ad}_{E_3} E_2$};
\node (mx4 y4) at (-4*\ra,1+4) {{\color{blue}$\hat W^{(1)}_{-m}$}};
\node (x0 y0) at (0,1) {$1$};
\draw [blue,thick,->] (mx1 y1) -- (mx2 y2);\draw [blue,thick,->] (mx2 y2) -- (mx3 y3);
\draw [red,thick,->] (mx1 y0) -- (mx1 y1);\draw [red,thick,->] (mx1 y1) -- (mx1 y2);\draw [red,thick,->] (mx1 y2) -- (mx1 y3);
\draw [blue,thick,->] (mx1 y2) to  (mx2 y4);
\draw [blue,thick,->] (mx1 y0) -- (mx2);
\draw [blue,thick,->] (mx2) -- (mx3);
\draw [blue,thick,loosely dotted, <-] (mx4 y4) to (mx3 y3);
\draw[red, thick, loosely dotted, ->] (mx1 y3) to (mx1 y4);
\draw[blue, thick, loosely dotted, ->] (mx3) to (mx4 y0);
\end{tikzpicture}
\end{center}


In the present paper, we provided an exhaustive description of the new integrable systems. It is done in two cases.

In the simplest case, we describe it in terms of matrix operators, which gives rise to integrable systems generalizing the rational Calogero model at the free fermion point, the eigenvalues of the matrix being associated with coordinates of particles of the Calogero system. Within each series (ray), the Hamiltonians at the free fermion point are just traces of powers of one and the same operator, which is a natural higher-order generalization of ${\p/\p\Lambda}$ for the Calogero model.

In the generic case, generalizing the Calogero system at generic coupling, they are associated with the WLZZ operators and are described by the $\beta$-deformation. In this case, the Hamiltonians are just sums of powers of one and the same operator, which is a natural higher-order generalization of the Dunkl operator $\hat D_i$ for the Calogero model.

We hope that the very important directions opened in \cite{ChR2} would develop fast,
and reveal many new properties of matrix models and integrable systems beyond their
traditional considerations.

\section*{Acknowledgements}

We are grateful to A. Popolitov for his interest and discussions. This work was supported by the Russian Science Foundation (Grant No.23-41-00049).


\begin{thebibliography}{12}

\bibitem{MSh}
A. Morozov and Sh. Shakirov, 
JHEP {\bf 04} (2009) 064, arXiv:0902.2627

\bibitem{Al} A. Alexandrov, Mod. Phys. Lett. {\bf A26} (2011) 2193-2199, arXiv:1009.4887

\bibitem{WLZZ1} R.~Wang, C.~H.~Zhang, F.~H.~Zhang and W.~Z.~Zhao,
Nucl. Phys. \textbf{B985} (2022) 115989,
arXiv:2203.14578

\bibitem{WLZZ2}
R. Wang, F. Liu, C.H. Zhang and W.Z. Zhao,
Eur. Phys. J. {\bf C82} (2022) 902, arXiv: 2206.13038

\bibitem{siMI} A.~Mironov and A.~Morozov,
Phys. Lett. \textbf{B835} (2022) 137573,
arXiv:2201.12917

\bibitem{ChR1} A.~Mironov, V.~Mishnyakov, A.~Morozov, A.~Popolitov, R.~Wang and W.~Z.~Zhao,
arXiv:2301.04107

\bibitem{toprec1} L. Chekhov and B. Eynard,
JHEP \textbf{0603} (2006) 014, hep-th/0504116

\bibitem{toprec2} L. Chekhov and B. Eynard,
JHEP \textbf{0612} (2006) 026, math-ph/0604014

\bibitem{toprec3} A. Alexandrov, A. Mironov and A. Morozov,
Theor. Math. Phys. \textbf{150} (2007) 153-164,
hep-th/0605171

\bibitem{toprec4} A. Alexandrov, A. Mironov and A. Morozov,
Physica {\bf D235} (2007) 126-167, hep-th/0608228

\bibitem{toprec5} N. Orantin,
arXiv:0808.0635

\bibitem{toprec6} A. Alexandrov, A. Mironov and A. Morozov,
JHEP {\bf 12} (2009) 053, arXiv:0906.3305

\bibitem{MMsc}  A.~Mironov and A.~Morozov,
arXiv:2210.09993, to appear in JHEP

\bibitem{ChR3} F.~Liu, A.~Mironov, V.~Mishnyakov, A.~Morozov, A.~Popolitov, R.~Wang and W.~Z.~Zhao,
arXiv:2303.00552

\bibitem{ChR4} L.~Y.~Wang, V.~Mishnyakov, A.~Popolitov, F.~Liu and R.~Wang,
arXiv:2301.12763

\bibitem{ChR2} A.~Mironov, V.~Mishnyakov, A.~Morozov, A.~Popolitov and W.~Z.~Zhao,
Phys. Lett. \textbf{B839} (2023) 137805,
arXiv:2301.11877

\bibitem{Hikami} K. Hikami and M. Wadati,
J. Phys. Soc. Jap. {\bf 62} (1993) 3857-3863

\bibitem{AM1} H.~Awata, Y.~Matsuo, S.~Odake and J.~Shiraishi,
Phys. Lett. \textbf{B347} (1995) 49-55,
hep-th/9411053

\bibitem{AM2} H.~Awata, Y.~Matsuo, S.~Odake and J.~Shiraishi,
Soryushiron Kenkyu \textbf{91} (1995) A69-A75,
hep-th/9503028

\bibitem{Wang} C.~H.~Zhang and R.~Wang,
Int. J. Mod. Phys. \textbf{A35} (2020) 2050137

\bibitem{Awata} H. Awata, {\sl Hidden Algebraic Structure of the Calogero-Sutherland Model, Integral Formula for Jack Polynomial and Their Relativistic Analog,} in: ``Calogero-Moser-Sutherland Models", CRM Series in Mathematical Physics, Jan Felipe van Diejen, Luc Vinet (eds.), Springer-Verlag New York, Year: 2000, pp. 23-35

\bibitem{Horozov1} E.~Horozov,
Ann. Inst. Fourier, Grenoble {\bf 55,} 6 (2005) 2069-2090

\bibitem{Horozov2} E.~Horozov,
Bulg. J. Phys. \textbf{36} (2009) 147-169

\bibitem{GJ} D. Goulden, D.M. Jackson and A. Vainshtein,
Ann. of Comb. {\bf 4} (2000) 27-46,
Brikh\"auser, math/9902125

\bibitem{MMN1}
A.~Mironov, A.~Morozov and S.~Natanzon,
Theor. Math. Phys. \textbf{166} (2011) 1-22,
arXiv:0904.4227

\bibitem{MMN2}
A.~Mironov, A.~Morozov and S.~Natanzon,
J. Geom. Phys. \textbf{62} (2012) 148-155,
arXiv:1012.0433

\bibitem{MMM} A.~Marshakov, A.~Mironov and A.~Morozov,
Phys. Lett. B \textbf{274} (1992) 280-288,
hep-th/9201011

\bibitem{SV} A.N. Sergeev and A.P. Veselov,
arXiv:0910.1984

\end{thebibliography}
\end{document}